\documentclass{article}

\usepackage{arxiv}
\usepackage[utf8]{inputenc} % allow utf-8 input
\usepackage[T1]{fontenc}    % use 8-bit T1 fonts
\usepackage{xcolor, hyperref, siunitx}
\usepackage[square,sort,comma,numbers]{natbib}

\usepackage{graphicx}
\usepackage{amssymb}
\usepackage{amsmath}
\usepackage{mathrsfs}

\DeclareSIUnit{\cells}{\text{cells}}

\title{Loop Current influence on the connectivity of the West Florida Shelf and impacts on red tide events}

\author{
  P. Miron\thanks{Email address for correspondence: pmiron@fsu.edu}\\
  Center for Ocean-Atmospheric Prediction Studies\\
  Florida State University, Tallahassee, FL, USA
}

\begin{document}

\maketitle

\begin{abstract}
We investigate the influence of the Loop Current (LC) on the connectivity of the Gulf of Mexico, with a focus on the West Florida Shelf (WFS), using in situ trajectories from satellite-tracked drifting buoys in the Gulf of Mexico (GoM). We subset the dataset into two groups, Loop Current extended and retracted phases, that are used to construct two Markov Chains representing the distinct underlying dynamics during those two periods. The LC phases were found to impact and modify substantively the general connectivity of the GoM. Additionally, we highlight the presence of almost-invariant regions, where particles tend to remain for an extended period, on the WFS when the LC is extended. Those regions are bounded by a previously identified Cross-Shelf Transport Barrier and correspond to records of high-density areas of Karenia Brevis, a dinoflagellate responsible for red tides. Finally, we show that Markov Chain modeling could help forecast harmful algae blooms by using the devastating 2017 red tide event as a test case. The development phase on the WFS and its further propagation to Florida's east coast ecosystems is presented.
\end{abstract}

\begin{keywords}
{Connectivity, Loop Current system, West Florida Shelf, Harmful marine algae, Red tides, Gulf of Mexico, Karenia brevis}
\end{keywords}

\section{Introduction}

The Loop Current (LC) is the dominant oceanic feature in the Gulf of Mexico (GoM), and carries warm water from the Caribbean Sea to the Atlantic Ocean. The LC system cycle can be summarized by three sequential events: the northern extension of the LC in the GoM, the shedding of a large anticyclonic eddy every \SIrange{3}{17}{}\,months, and the retraction of the LC. Although seasonality can have impacts on the coastal circulation of the GoM \citep{Zavala-Hidalgo-2003}, the large-scale circulation is dictated by the LC system cycle.

Over the past decades, the study of the Gulf of Mexico's (GoM) connectivity was made possible due to the development of remote sensing techniques and the increasing number of Lagrangian datasets of surface buoys and subsurface floats. In \cite{Miron-etal-2017} and \cite{Miron-etal-2019}, we respectively described the connectivity of the surface and deep GoM using satellite-tracked surface buoys and subsurface acoustically-tracked floats. Based on the combination of the past 30 years of publicly available observations, the \textit{averaged} connectivities (surface and depth) were presented. Temporal analysis (monthly or seasonal) of the GoM connectivity is challenging because the circulation is largely controlled by the non-periodic LC. To further deepen connectivity analyses in the GoM, as part of this study, we propose to evaluate the influence of the LC on the connectivity of the GoM by splitting the historical dataset into two subsets during the extended and retracted phases. More precisely, we will focus on the influence of the LC phases on the West Florida Shelf (WFS) and their impacts on harmful algae bloom (HAB) in the area.

Commonly referred to as red tides, HABs in the GoM are caused by a single-celled organism named Karenia brevis (K. brevis). Red tides are found in several locations across the GoM, such as the Texas shelf \citep{Magana-etal-2003, Hetland-Campbell-2007}, and the Yucatan coastal zone \citep{Enriquez-etal-2010}, with the largest blooms occurring on the WFS \citep{Walsh-etal-2006}. Red tide occurs almost annually on the WFS, and has devastated effects on the ecosystems, fisheries, tourism industry, and even human health \citep{Flewelling-etal-2005, Landsberg-2002, Shumway-etal-2003}. Because of these consequences, K. brevis has been the focus of many studies over the past seventy-five years. We refer the reader to seminal works \citep{Steidinger-1975, Steidinger-Haddad-1981} and more recent analyses on their proliferation \citep{Walsh-etal-2001, Walsh-etal-2002, Walsh-etal-2003, Weisberg-etal-2019}. Although the life cycle of K. brevis is well known, the various external factors required to support red tide, understand their propagation, and forecast large events remain unclear.

The main goals of this study are to evaluate the impact of the LC cycle on the connectivity of the GoM, and analyze how circulation and connectivity variations on the WFS can impact the concentration of HABs in this region. The publication goes as follows. In Sec.~\ref{sec:methods}, we present the selected data source, the Markov Chain developed framework, and the concept of dynamical Lagrangian geography. In Sec.~\ref{sec:results}, we present the two models associated with the extended and retracted states of the LC, and evaluate their influence on the connectivity of the GoM, with an emphasis on the impact on the WFS. This is followed by a discussion in Sec.~\ref{sec:discussion} on the implications of HABs development on the WFS. Finally, a conclusion is presented in Sec.~\ref{sec:conclusion}.

\section{Material and methods}\label{sec:methods}

\subsection{Lagrangian observations}\label{sec:observations}

To obtain an approximation of the ocean's surface dynamics, we used all publicly available drifter trajectories in the GoM between 1992 and 2020 \citep{Lilly-Perez-Brunius-2021}. This dataset contains 2,731 freely available satellite-tracked (Argos and GPS) surface drifter trajectories from 15 different sources. It includes a wide range of drifter types, with various dimensions and designs, that samples the area at several temporal frequencies. All drifters are attached to a drogue to limit the influence of winds and waves. The type and size of drogues vary and are centered at various depths between \SIrange{1}{45}{\meter}. For a complete description of the data sources and experiments, the reader is referred to \cite{Lilly-Perez-Brunius-2021a}.

\subsection{Markov Chain}\label{sec:theory}

A Markov chain is a stochastic model that forecasts the future state of a system based strictly on its current state. To create the model, we assume that historical drifter trajectories are a reliable proxy of the surface ocean dynamics and that they are advected by a time-homogeneous stochastic process (i.e., steady advection-diffusion).

The construction of the transition matrix, which represents the underlying Markov process, involves a series of steps. First, the domain $D$ is divided into $N$ boxes ${b_i}_{\,i \in S}$, with $S = \lbrace 1,2, \dots, N\rbrace$. Second, the drifter trajectories are split into $T$ day-long segments; each segment $i$ is formed by the coordinates $x^i_t$ and $x^i_{t+T}$, and the complete set of observations are noted $X_t$ and $X_{t+T}$. Third, the transition matrix $P$ is constructed by evaluating the conditional transition probabilities between boxes of the domain,
\begin{equation}
    P_{ij} = \Pr(X_{t+T} \in b_j | X_t \in b_i) \;\text{with}\ i,j \in S
    \label{eq:markovchain}
\end{equation}
where coefficients of $P_{ij}$ represent the transitional probability between $b_i$ to $b_j$ during $T$ days. Here, we set $T=5$ days, which is large enough to allow interbox connections and have minimal correlation with past positions \citep[see][for more details]{Miron-etal-2017}. In practice, \eqref{eq:markovchain} is approximated by counting the transitions of the observations, 
\begin{equation}
    P_{ij} = \frac{C_{ij}}{\sum_{k\in S} C_{ik}},\;\text{with}\; C_{ij} = \lbrace X_t \in b_i, X_{t+T} \in b_j\rbrace,
    \label{eq:transitionmatrix}
\end{equation}
where $C_{ij}$ is the number of observations in $b_i$ at any time $t$ and in $b_j$ at $t+T$. Because each row of $P$ in \eqref{eq:transitionmatrix} is normalized, $P_{ij}$ is row-stochastic (i.e. $\sum_{j\in S} P_{ij} = 1$).

Once constructed, the transition matrix $P$ can be used to forecast the temporal evolution of a discrete probability distribution $\bf f = (f_1, f_2, \dots, f_N)$, such as an initial distribution of harmful algae across $D$. The evolved distribution represents the spread of the initial tracer probability. The evolution of such distribution after $T, 2T, \dots, kT$ days is obtained by repeatedly left-multiplying using the transition matrix,
\begin{align}
    f_{t + T} &= f_t P \nonumber\\
    f_{t + 2T} &= (f_tP)P = f_t P^2\nonumber\\
    f_{t + kT} &= f_t P^k \;\text{for any integer}\ k\ge 1.
    \label{eq:pushforward}
\end{align}

Note that since the distribution is initially normalized and $P$ is row-stochastic, the sum of the evolved distribution is always conserved and equal to $1$ (mass conservation). Due to the probabilistic nature of the Markov Chain, the obtained forecasts are based on the complete variability over the 30 years of the observations, as opposed to a forecast based on mean Eulerian flow fields from the same initial observations.

For more details on the theory, the reader is referred to a previous publication \citep{Miron-etal-2019}. A python module is also publicly available to reproduce the presented results or to analyze other datasets or regions \citep{Miron-Helfmann-2021}.

\subsection{Dynamical Lagrangian geography}

From the eigenspectrum of the transition matrix \eqref{eq:transitionmatrix}, it is possible to reveal areas where trajectories (or a tracer) tend to stay and accumulate for a long period of time. Those areas are key in assessing the connectivity of a domain. Corresponding to \emph{almost}-invariant sets, these areas are extracted from the inspection of the eigenvectors of $P$ with eigenvalues $\lambda$ close to 1. More precisely, the structures of the left eigenvectors correspond to regions of attraction (or \emph{almost}-invariant sets), while the structures depicted by the right eigenvectors correspond to their corresponding basins of attraction \citep{Froyland-etal-2014}.

The magnitude of the eigenvalues $\lambda_i$ associated with the eigenvectors quantifies the residence time of the region, where longer-lasting and \textit{robust} sets have $\lambda \approx 1$. The basins of attraction are identified as the constant region of the right eigenvectors associated with the largest eigenvalues. The decomposition of the domain into disjoint basins of attraction was first introduced in \cite{Froyland-etal-2014} and is now known as the dynamical Lagrangian geography. The regions of a \textit{geography} are disconnected and allow direct assessment of the connectivity. Furthermore, the boundaries delimiting those regions are directly determined by the underlying Lagrangian circulation.

In previous publications, the regions of the Lagrangian geographies were manually extracted from the top right eigenvectors and then stitched together. To automate this process and make results reproducible, we propose to use the Sparse EigenBasis Approximation (SEBA) algorithm. Presented in \cite{Froyland-etal-2019}, it simplifies spectral clustering by automatically disentangling features---regions of the dynamical Lagrangian geography---from a given list of eigenvectors. Another advantage is that the algorithm does not require setting a priori the number of regions (which is not trivial in practice), such as k-means and other clustering methods.

\section{Results}\label{sec:results}

\begin{figure*}
    \centering
    \includegraphics[width=\textwidth]{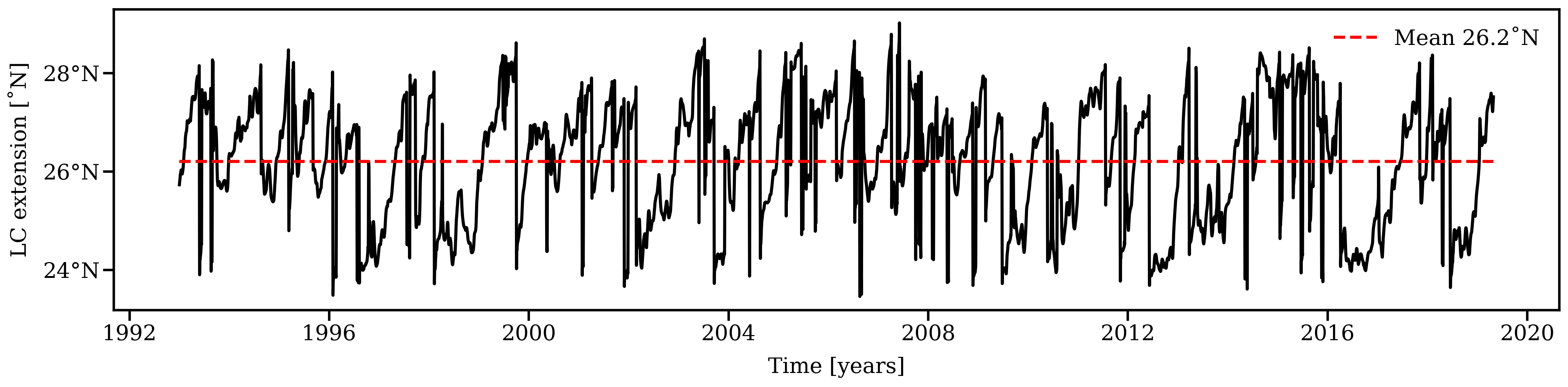}
    \caption{Time series of the northern LC extension (black line) for the complete period between 1992--2020. The mean extension value is shown by the red dashed line.}
    \label{fig:lc-timeseries}
\end{figure*}

As part of this study to evaluate the impact of the LC on the GoM's connectivity, we split the Lagrangian dataset into two subsets associated with the two main phases---extended and retracted---of the LC. The phases are defined as a function of the \SI{17}{\centi\meter} contour of sea surface height (SSH), which is extracted following the methodology described in \cite{Hiron-etal-2020}. This SSH contour is commonly used to identify the LC boundary. The time series of the northern extension of the Loop Current (black), taken as the maximum latitude of the \SI{17}{\centi\meter} contour line inside the GoM, is shown in Fig.~\ref{fig:lc-timeseries} along with the average extension (dashed red) at \SI{26.2}{\degree}N over complete time series. From this time series, trajectories are grouped into segments occurring during the extended phase (above the mean value) and the retracted phase (below the mean value), creating two subsets associated with both LC phases.

From those two datasets, two Markov chains are created; each transition matrix representing the dynamics of the ocean surface during one of the LC phases. The creation of both Markov chains involves a few steps, which are repeated for each subset of trajectories corresponding to both phases of the LC. First, we interpolate daily the drifter trajectory segments described in Sec.~\ref{sec:observations}. Second, we form two arrays: one containing the positions at any instant during the 1992--2020 period ($x^i_t$) and another with the corresponding position ($x^i_{t+T}$) after $T=\SI{5}{\day}$. Third, we define a grid covering the region with boxes of \SI{0.5}{\degree} width or about \SI{3025}{\kilo\meter\squared}. For the current domain, which includes the GoM and part of the Caribbean Sea north of \SI{18}{\degree}N, there are a total of $770$ boxes in the grid. Ultimately, using the two arrays previously created, the coefficients of the transition matrix $P$ are obtained using \eqref{eq:transitionmatrix} by evaluating the transitional probabilities between boxes of the domain.

\begin{figure*}
    \centering
    \includegraphics[width=\textwidth]{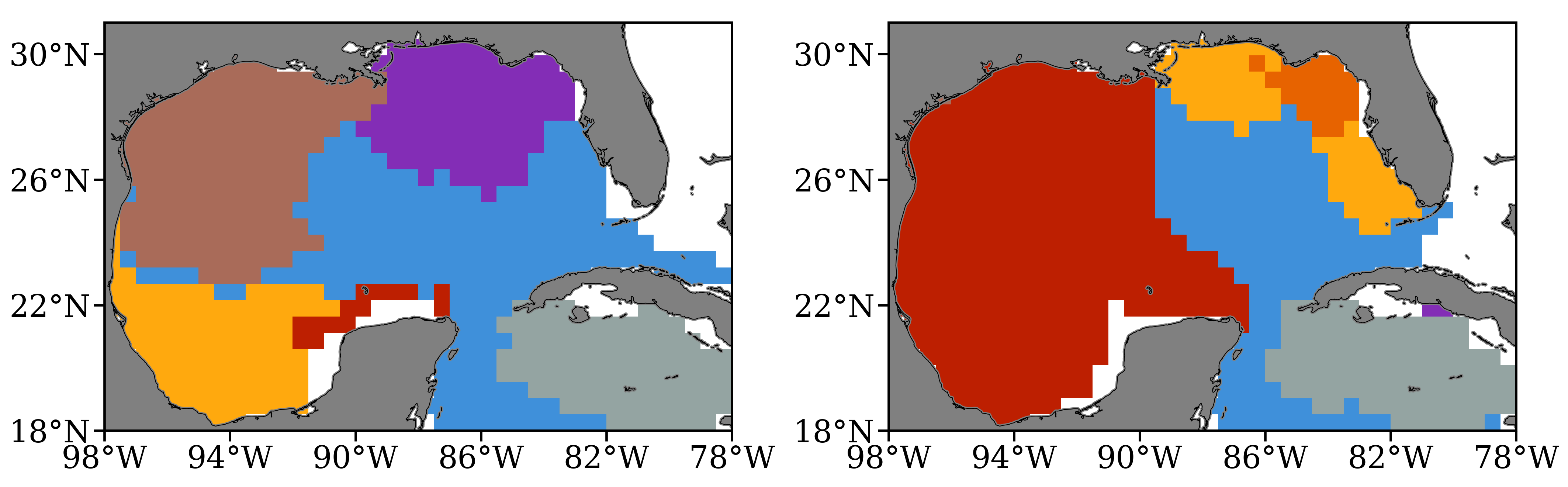}
    \caption{The Lagrangian geography of the GoM for the retracted (left) and extended (right) phase of the LC.}
    \label{fig:regions}
\end{figure*}

The Lagrangian geographies, presented in Fig.~\ref{fig:regions}, were obtained by combining the six right eigenvectors associated with the largest eigenvalues (with $\lambda \gtrapprox 0.9$) using the SEBA algorithm \citep{Froyland-etal-2019} on the retracted (left) and extended (right) transition matrices. Recall that the structures of the right eigenvectors are associated with basins of attraction of different attractors, which implies that regions of Lagrangian geographies are dynamically disconnected from each other. The number of eigenvectors was selected similarly to previous studies to facilitate comparisons, although the main result is still observed using the top two dominant eigenvectors. 

During both LC phases, we observe a region (gray) south of Cuba that extends from \SIrange{78}{86}{\degree}W, which is characterized by the recurrent presence of large cyclonic gyres \cite{Centurioni-etal-2003}. In addition, the central region (blue) highlights the location of the LC and its direct connection from the Yucatan channel to the Straits of Florida. During the retracted phase (left), that region extends westward of \SI{90}{\degree}W due to the presence of large LC anticyclonic eddies propagating westward during that phase. During the extended phase (right), the blue region extends northward to \SI{28}{\degree}N in the eastern GoM analogously to the LC. Other regions of the geographies differ considerably as a function of the LC phases. We note the differences in the west GoM, where the northern and southern regions (brown and yellow) are disconnected during the retracted phase. Also during this phase, a small region (red) covers the Yucatan shelf. As expected, those dynamical Lagrangian geographies are different from the geography obtained using the complete dataset \citep[right panel of Fig. 6 of][]{Miron-etal-2017}. As part of this study, we turn our focus on the impact of the LC, which isolates the WFS region (yellow and orange) during an extended phase, and have important impacts on the region's connectivity.

\begin{figure}
    \centering
    \includegraphics{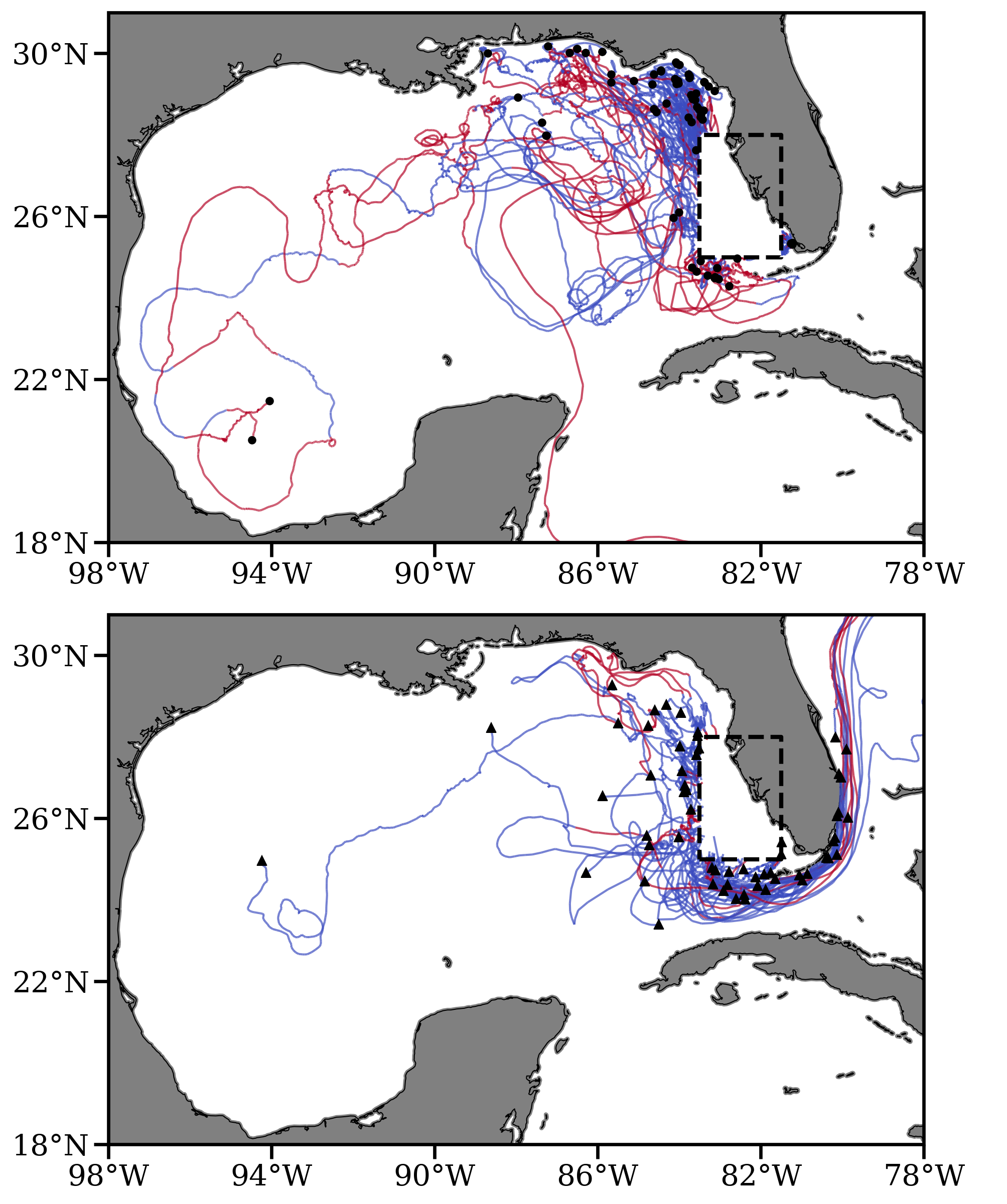}
    \caption{Trajectories of drifters reaching (top) and leaving (bottom) the region delimited by the black rectangle. For both panels, blue segments correspond to a period of retracted LC, while red segments correspond to a period of extended LC. The black scatters (circle and triangle) show the launch locations (top) and the end locations (bottom) of the trajectories. Trajectories inside the rectangle region are not shown for clarity.}
    \label{fig:wfsdrifters}
\end{figure}

To directly observe the influence of the LC on the connectivity, trajectories reaching and leaving the area delimited by the black dashed rectangle are respectively shown on the top and bottom panels in Fig.~\ref{fig:wfsdrifters}. Although more isolated than other regions of the GoM, the WFS is sampled by a total of 152 drifters. In Fig.~\ref{fig:wfsdrifters}, trajectory segments are colored according to the LC phase, with the retracted phase in blue and the extended phase in red. In the top panel, there are 100 trajectories entering the region. From its western edge, drifters reach the area during both phases of the LC. In contrast, from Homosassa Bay to Apalachee Bay (northern edge) and from the Dry Tortugas National Park (southern edge), the connections occur during a specific phase of the LC. More precisely, during retracted phases, most drifters reach the area from the north, whereas during extended phases, a few drifters enter the area and are mostly from the south. Although, this might be biased by important drifters deployment near the Dry Tortugas National Park, as shown by the black circle scatters.

In the bottom panel, the majority of the 83 connections out of the area occur during the LC retracted phase (blue). Once in the area, 45\% of the trajectories are trapped and terminated in the region; which corresponds to 69 drifters including 40 deployed inside the area. Furthermore, the majority of the trajectories exiting the region terminate just south of the WFS and along the Florida Keys, as shown by the black triangle scatters. Some drifters exiting during the extended phase were attracted by strong LC frontal eddies, which are able to attract particles located near the WFS break, as presented in a recent study by \cite{Hiron-etal-2022}.

\section{Discussion}\label{sec:discussion}

The influence of the LC on the connectivity of the WFS and its isolation from the eastern GoM was presented in Sec.~\ref{sec:results}. In this section, we discuss how the connectivity of the WFS can impact HABs, commonly referred to as red tides on the WFS. Important to note that Florida red tides are not uncommon; they occur annually, mostly in the region of Tampa Bay and Charlotte Harbor, and were first documented back in 1844. Because of their major impacts on tourism, fisheries, ecosystems, and even human health, red tides have been the focus of multiple studies \citep{Olascoaga-etal-2006, Olascoaga-2010, Maze-etal-2015} and are constantly monitored by the Florida Fish and Wildlife Conservation Commission (FWC). 

Florida red tides are caused by K. brevis---a dinoflagellate---that produces neurotoxins that cause respiratory problems in humans and can attack the nervous systems of fish, mammals, and other wildlife \citep{Steidinger-Haddad-1981}. To thrive, dinoflagellates require light, adequate water temperature and salinity, and nutrients (nitrogen and phosphorus). GoM waters are commonly considered oligotrophic, with levels of inorganic nitrogen and phosphorus too low to support HAB; a nutrient input is mandatory to support red tide blooms. 

Various sources of nutrients are listed in the literature. The first source of nutrients comes from river outflows, which are also responsible for creating an ideal stratified area where dinoflagellates can out-compete other planktons \citep{Maze-etal-2015}.
The second source comes from the deep GoM and is brought to the continental shelf through upwelling processes \citep{Tester-Steidinger-1997, Lanerolle-etal-2006, Weisberg-Liu-2022}. Finally, large HABs also appear to be linked to aeolian supply of nutrients \citep{Walsh-Steidinger-2001}, but those constitute minor contributions and are insufficient to support large blooms \citep{Vargo-etal-2008}. Single-celled eukaryotes can survive on different sources of nutrients, so it is not trivial to pinpoint the specific source of nitrogen or phosphorus.

\begin{figure*}
    \centering
    \includegraphics[width=\textwidth]{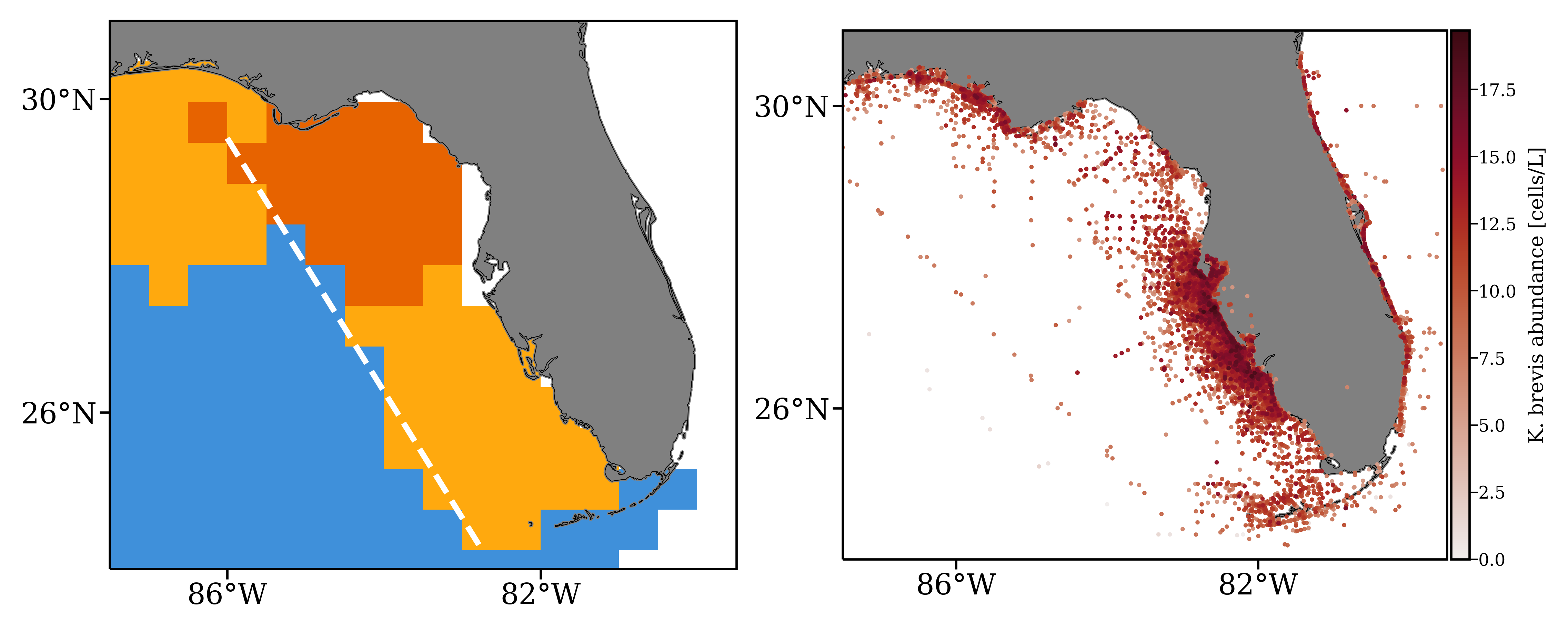}
    \caption{Comparison of the density of K. brevis on the West Florida shelf (right) with the Lagrangian geography of the area during an extended LC phase (left). The white dashed line is the cross-shelf transport barrier. The K. brevis concentration dataset between 1954 to 2020 was obtained from Florida Fish and Wildlife Research Institute.}
    \label{fig:density}
\end{figure*}

Described recently in \cite{Weisberg-Liu-2022}, red tides initiation zone is located along the \SI{50}{\meter} isobaths or about \SIrange{18}{74}{\kilo\meter} offshore \citep{Steidinger-1975} and are later transported eastward along the coast, where higher concentration are found \citep{Brand-Compton-2007}, shown in the right panel of Fig.~\ref{fig:density}. Once initiated, it becomes nearly self-sustained using biomass from the fish kill. It is interesting to compare the spatial distribution of historical concentrations of K. brevis with the Lagrangian geography presented in Fig.~\ref{fig:regions}. During the extended phase, the LC isolates the WFS from the rest of the GoM (zoom of the area in the left panel of Fig.~\ref{fig:density}). In fact, the WFS region (yellow) extends from the Mississippi River Delta to the Florida Keys, is confined closer to shore, and subdivided by a smaller region (orange) in the northeastern GoM where lower historical concentration levels are found, which contrasts dramatically with its counterpart (purple) during the retracted phase (left panel of Fig.\ref{fig:regions}).

In \cite{Maze-etal-2015}, the authors presented an analysis of the environmental conditions during large blooming events (define as periods with concentration higher than \SI{1e5}{\cells\per\liter}). They tested how factors such as the extension of the LC, wind patterns (and upwelling), seasonality, and river discharges, correlate with periods of large blooming. It was found that blooms occurred independently of seasons, with roughly 60\% occurring in winter and spring and 40\% in summer and fall. Similarly, wind patterns in the area show no significant correlation with blooming events. In summer, the south and southeast winds generate downwelling, while winds in winter are from the northeast and create important upwelling along the WFS \citep{Weisberg-etal-2001}. This suggests that upwelling alone is not sufficient to sustain important blooming events. Similarly, \cite{Maze-etal-2015} did not identify a clear link between fluctuations of river discharge and the presence of red tide. The same results were also observed by \cite{Dixon-Steidinger-2004}, who only identified correlations with a limited number of rivers across the WFS.

However, \cite{Maze-etal-2015} identified that the occurrence of HABs is statistically dependent on the position of the LC. In fact, all periods of large blooms occur during the extended phase of the LC, which as we showed, creates a region of enhanced retention close to the WFS. The cross-shelf barrier (white dashed line in the left panel of Fig.~\ref{fig:density}) was shown to restricts the circulation of the south WFS \citep{Yang-etal-1999, Olascoaga-etal-2006, Olascoaga-etal-2008}, creating a region denoted the Forbidden Zone south of the WFS, which was not visited by surface drifters. Using velocity fields from HF Radar and the HYbrid Coordinate Ocean Model (HYCOM), \cite{Olascoaga-2010} highlighted the persistence of the cross-shelf boundary across seasons, noting a translation closer to the coast during summer. Their three-dimensional analysis also showed that the barrier extends to the sea floor, preventing nutrients from leaving (and entering) the area across the complete water column. The presence of the cross-shelf transport boundary during the extended phase of the LC could help retains nutrients from the diverse sources previously described and create a suitable location for the K. brevis development. 

\begin{figure*}
    \centering
    \includegraphics[width=\textwidth]{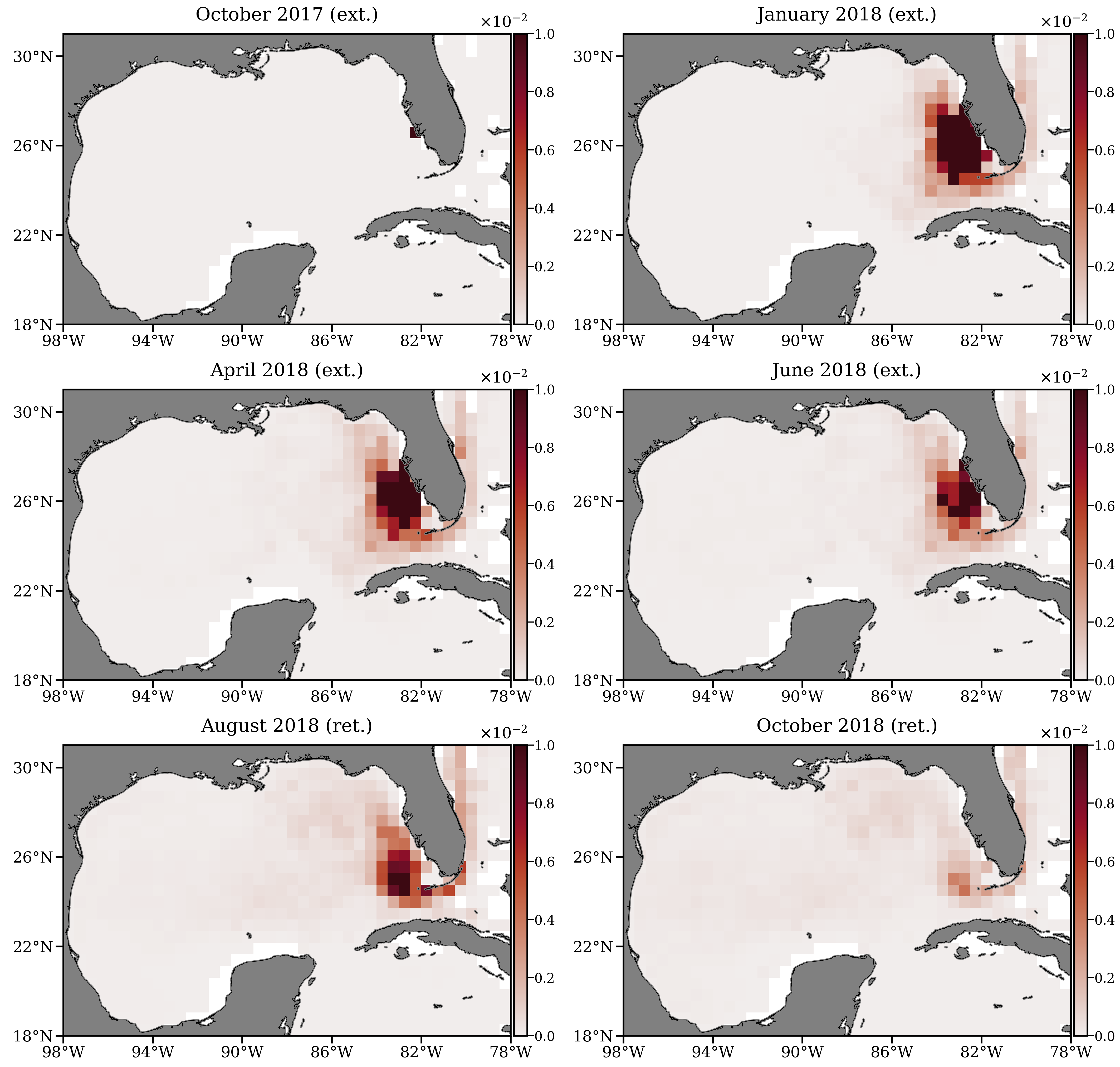}
    \caption{Evolution of a probability density initialized on October 2017 offshore the St. Petersburg's area (top-left panel). The other panels present the dispersion of the probability density during the devastating 2017--2019 red tide events. The date and the phase (extended and retracted) of the LC are indicated in the title of each panel.}
    \label{fig:evolution_tracers}
\end{figure*}

As a study case, we present in Fig.~\ref{fig:evolution_tracers} the evolution of a probability density, simulating the devastating red tide event on the WFS between October 2017 and January 2019. This event was one of the most intense in recent years, with a total economic impact of \$318 million \citep{Ferreira-etal-2022}. On the first panel of Fig.~\ref{fig:evolution_tracers}, the distribution is initialized offshore of the St. Petersburg's area (where high K. brevis concentrations are usually observed) at the beginning of the HAB event in October 2017. At the time, the LC was extended, so the distribution is initially evolved with \eqref{eq:pushforward} using the \textit{extended} transition matrix. In April 2018, LC eddy Revelle separated and reattached a few weeks later. At that time, in the third panel of Fig.~\ref{fig:evolution_tracers}, most 
of the probability distribution is on the WFS, remaining mostly inside the southern section of the yellow region highlighted in the Lagrangian geography associated with the extended phase of the LC (right of Fig.~\ref{fig:regions}). In August 2018, the LC eddy separated once more, and the LC retracted. From that time, the distribution is evolved using the \textit{retracted} transition matrix. With the retraction of the LC, we expect a faster transport from the WFS to the East Florida shelf (EFS) (left of Fig.~\ref{fig:regions}). Two and four months later, we observe this rapid propagation to the EFS on the last two panels of Fig.~\ref{fig:evolution_tracers}, and most of the probability distribution left the WFS. As seen with the trajectories out of the WFS (bottom of Fig.~\ref{fig:wfsdrifters}), most of the connections out of the area are through the southern edge of the shelf towards the Florida Keys. Those results align with the observations at the time of red tide reaching the Biscayne Bay (Miami area) and the West Palm Beach area in October 2018, a few weeks after the retraction of the LC.

\section{Conclusion}\label{sec:conclusion}

The main goals of this study were to evaluate the impact of the LC cycle on the large-scale connectivity of the GoM, and analyze how circulation and connectivity variations on the WFS can impact the concentration of HABs in this region. For that, we grouped drifter trajectories as a function of the phases of the LC to create the retracted and extended transition matrices. From the eigenspectrum of those matrices and by producing two Lagrangian geographies, we highlighted that the LC can modify large-scale connectivity of the GoM, and that the WFS becomes completely isolated from the rest of the GoM during extended LC phases.

\cite{Maze-etal-2015} showed a correlation between red tide events on the WFS and the extended phase of the LC. As part of this study, we identified regions of recirculation on the WFS, only present during extended LC periods, that coincide with areas of high concentration of K. brevis from historical data. The isolation of the WFS during extended phases can trap nutrients from river outflows, offshore upwelling, and diverse aerial sources, and create a suitable region for the development of HABs. It is also important to note that other conditions such as high temperature, sufficient nutrient concentration, and stratification are required for supporting large red tide blooms. Finally, we showed how time-dependent Markov Chain modeling can be useful in forecasting blooming events in the GoM by reproducing the 2017--2019 infamous red tide event.

The input of nutrients from rivers remains an open question. Although rivers outflow was not directly linked to the presence of red tides, \cite{Maze-etal-2015} highlighted seasonal influence with lower flows from several rivers during winter blooms than summer blooms. This could be explained by a seasonal variation of the nutrient concentration from the river outflows. To answer this question, nutrient concentrations should be obtained from river discharge at multiple locations along the WFS, alongside an approximation of nearshore nutrients input from other sources such as air pollution, fertilizer runoff, septic tanks, and wastewater systems.

\section*{Conflict of Interest Statement} 

The author declares that the research was conducted in the absence of any commercial or financial relationships that could be construed as a potential conflict of interest.

\section*{Funding}

This study was funded by the National Academy of Sciences, Engineering and Medicine (Gulf Research Program UGOS \#2000011056) and the Gulf Research Program of the National Academies of Sciences, Engineering, and Medicine under award \#2000013149. The content is solely the responsibility of the authors and does not necessarily represent the official views of the Gulf Research Program or the National Academies of Sciences, Engineering, and Medicine.
 
\section*{Acknowledgments}

I thank Josefina M. Olascoaga for discussions on red tide events during my time at the Rosenstiel School at the University of Miami. I also thank Luna Hiron for the discussions on the influence of the Loop Current System on the circulation of the WFS and the many suggestions to improve the manuscript.

\section*{Data Availability Statement}

The data supporting this study's findings are openly available in the NOAA Global Drifter Program dataset, available at https://www.aoml.noaa.gov/phod/gdp and in GulfDrifters: A consolidated surface drifter dataset for the Gulf of Mexico \citep{Lilly-Perez-Brunius-2021}. Finally, the numerical code \citep{Miron-Helfmann-2021}, and a Jupyter Notebook to reproduce the results are publicly available on GitHub (\href{https://github.com/philippemiron/pygtm/}{Code}, \href{https://github.com/philippemiron/pygtm/papers/redtide.ipynb}{Notebook}). The historical dataset (1954--2020) of harmful algal bloom can be obtained upon request from the Florida Fish and Wildlife Conservation Commission (FWC) \href{https://myfwc.com/research/redtide/monitoring/database/}{https://myfwc.com/research/redtide/monitoring/database/}.

\bibliographystyle{Frontiers-Harvard}
\bibliography{redtide-lc}

\end{document}